\documentclass[conference,compsocconf]{IEEEtran}
\usepackage{times}
\usepackage{graphicx, pstricks, algorithm, algorithmic}
\usepackage{subfig}
\usepackage{epstopdf}
\usepackage{balance}
\usepackage{multirow}
\usepackage{float}
\usepackage{bigstrut}
\usepackage{amsmath,amsfonts}
\usepackage{inputenc}
\begin{document}
%\tableofcontents
\IEEEoverridecommandlockouts

\title { Accelerating the Depth Reconstruction Algorithm with CUDA/GPU }

%\tableofcontents

\author{\IEEEauthorblockN{Ke Yue, Schwarz Nicholas, and Tischler Jonathan Z.\thanks{The research is supported
      by ANL .}}
   \IEEEauthorblockA{Department of APS\\
     Argonne National Lab\\
     Argonne, Illinois 60439\\
     Email: \{kyue, nschwarz, tischler\}@aps.anl.gov}
}

\maketitle

\begin{abstract}
The Laue diffraction microscopy uses the polychromatic Laue micro-diffraction technique to examine the structure of materials with sub-micron spatial resolution in all three dimensions. Properties that can be measured include local crystallographic orientations, orientation gradients and strains. All those data is recorded in HDF5 format \cite{folk1999hdf5} and those images will be processed with a depth reconstruction algorithm on CPU. In recent years, GPU (Graphics processing units) on commodity video cards is well known for its powerful computational capability. In this paper, we propose a GPU program solution on the depth reconstruction problem instead of running the program on CPU. The results showed that the test running time would be 25\% to 30\% of the prior CPU design.

\end{abstract}

\IEEEpeerreviewmaketitle

 \section{Introduction}
 \label{sec:introduction}
 The Laue diffraction microscopy~\cite{larson2002three} \cite{liu2011x} uses the polychromatic Laue micro-diffraction technique to examine the structure of materials with sub-micron spatial resolution in all three dimensions. The materials which are investigated include inter- and intra-granular orientation distributions in polycrystals ,elastic strain tensors in elastically deformed materials, and plastic deformation
microstructures under microindents in Cu single crystals. This structural microscopy techniques is very powerful for detailed investigation of the microstructure and evolution in materials, especially including local crystallographic orientations, orientation gradients and strains. The data set get from the structural microscopy technique used in sector 34ID at Advanced Photon Source of Argonne National Laboratory is recorded in HDF5 format and processed by a depth reconstruction program running on CPU.

 GPU  (graphics processing units) is a poweful computational device for 3D graphics processing. In 2006, NVIDIA releases a new GPU architecture which facilities efficient general purpose computing on GPU (GPGPU). Unlike CPU, however, GPU has a parallel throughput architecture that emphasizes executing many concurrent threads slowly, rather than executing a single thread quickly. GPU with massively parallel structure is more efficient than the general purpose CPU~\cite{che2008performance} for processing large blocks of data in parallel.

\begin{figure}
    %\centering
    \includegraphics[width = 0.8\linewidth]{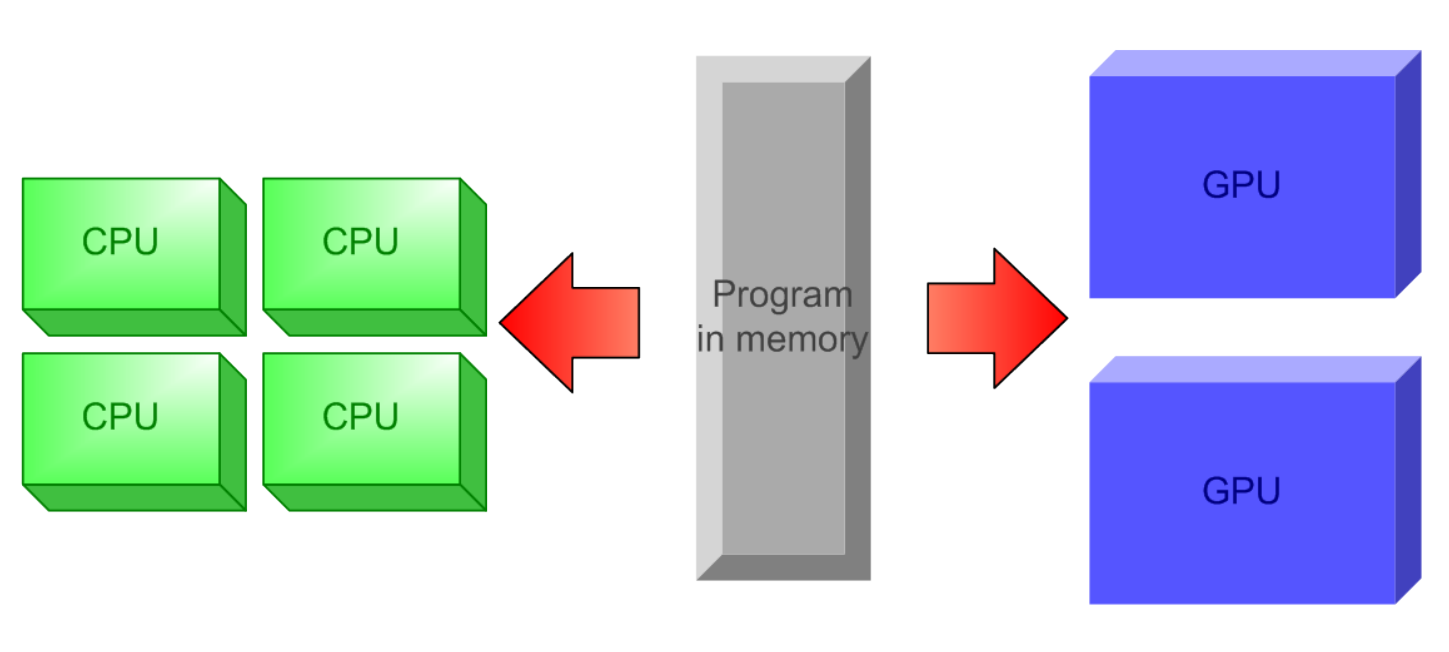}
    \caption{GPU program}
    \label{fig:cpugpu}
\end{figure}

CUDA (Compute Unified Device Architecture) is a parallel computing platform and programming model created by NVIDIA in 2007 \cite{nvidia2008programming} and implemented by the GPU. CUDA gives the developers access to the virtual instruction set and memory of the parallel computational elements in GPUs via C programming environment. Using CUDA, the latest NVIDIA GPUs become accessible for computation tasks like CPUs as showed in Fig.~\ref{fig:cpugpu}. And GPU program using CUDA has been used widely used on scientific problems such as~\cite{shalaby2013using},~\cite{hobson2013challenges},~\cite{choe2013gpu} and promotes the total program running speed.

This paper proposes a new CUDA implementation for depth reconstruction algorithm of HDF5 images. The result shows that the new CUDA design only needs 25\% to 30\% of the original CPU design's running time. Section~\ref{sec:algorithm} describes the design issues for the depth reconstruction algorithm and CUDA solution to the problem.  Section~\ref{sec:experiment} shows the experiment result and the comparison between the original CPU design and the new CUDA design.

 \section{Related Work}
 \label{sec:related}
 It has been mentioned in several publications that transferring s significant amount of data between CPU and GPU through PCI Express link can cause a throughput bottleneck due to the limited bandwidh of PCI Express link. Schaa and Kaeli~\cite{schaa2009exploring} admits that PCI Express will be a throughput bottleneck if a full dataset can not fit into the memory on a GPU. Owens et al.~\cite{owens2008gpu}, Fan et al.~\cite{fan2004gpu}, Cohen and Molemaker~\cite{cohen2009fast} and Dotzler et al.~\cite{dotzler2010jcudamp} all showed same concerns and suggest that, in the cuda program, the communication between CPU and GPU through PCI Express should be as least as possible.

Komoda et al. propose a library for OpenCL that
automatically overlaps computation with data communication
given the memory usage pattern of a kernel~\cite{komoda2012communication}. The
performance of the resulting applications is close to that of
the double-buffering scheme. However, the programmer is still
required to provide various details on the data usage pattern
of the kernel.
CGCM and DyManD are two systems that automate
CPU-GPU data communications through a hybrid
compile-time and run-time scheme~\cite{jablin2012dynamically},~\cite{jablin2011automatic}. However,
the data transfers are not overlapped with computation. And
the programmer is still responsible for splitting the data into
smaller chunks and invoking the kernel multiple times.
Pai et al. propose a system that automates CPU-GPU
memory management based on a coherence scheme in order
to reduce superfluous communication~\cite{pai2012fast}. To do this, when
a data item is accessed on one side (CPU or GPU side), it is
transferred (from the other side) if it is not locally available
or if its local version is stale. This system does not overlap
computation and communication.

\section{Cuda Solution for Reconstruction Program}
\label{sec:algorithm}
\subsection{Problem}
\label{subsec:problem}

\begin{figure}
    %\centering
    \includegraphics[width = 0.95\linewidth]{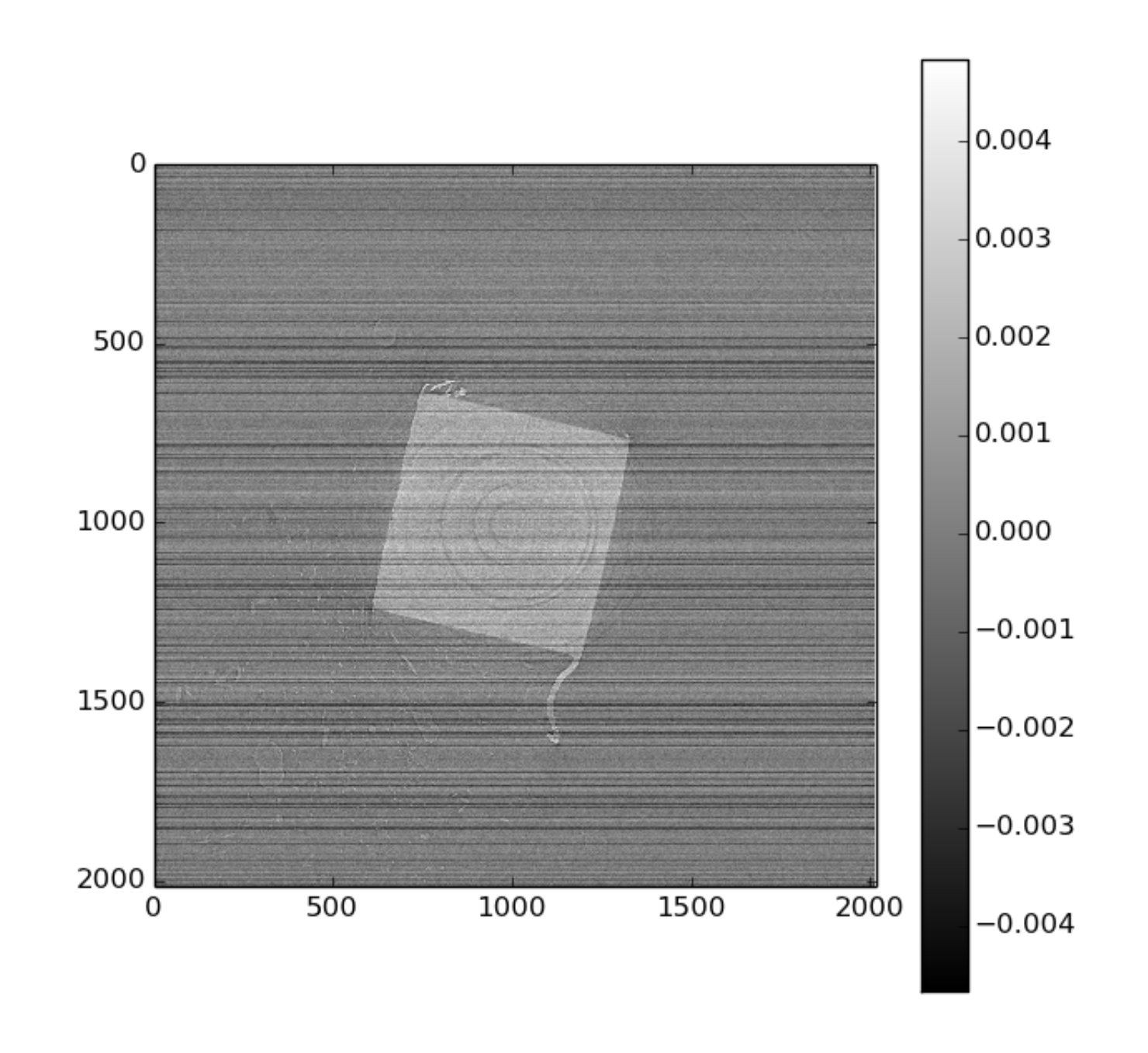}
    \caption{HDF5 Image}
    \label{fig:hdf5}
\end{figure}

The prior design which uses CPU for the image depth reconstruction takes long time for handling big size data set. Thus we need to figure a more efficient way for processing huge amount of data. In the following section we will propose a GPU program design which could improve the efficiency of image depth reconstruction. The input data would be a set of 2D images with pixel intensity values saved in HDF5 format~\cite{folk1999hdf5}. Fig.~\ref{fig:hdf5} shows a sample HDF5 image used as the data input.

\subsection{Issues for Cuda Program Design}
\label{subsec:issue}

\begin{figure*}
    \centering
    \includegraphics[width = 1.0\linewidth]{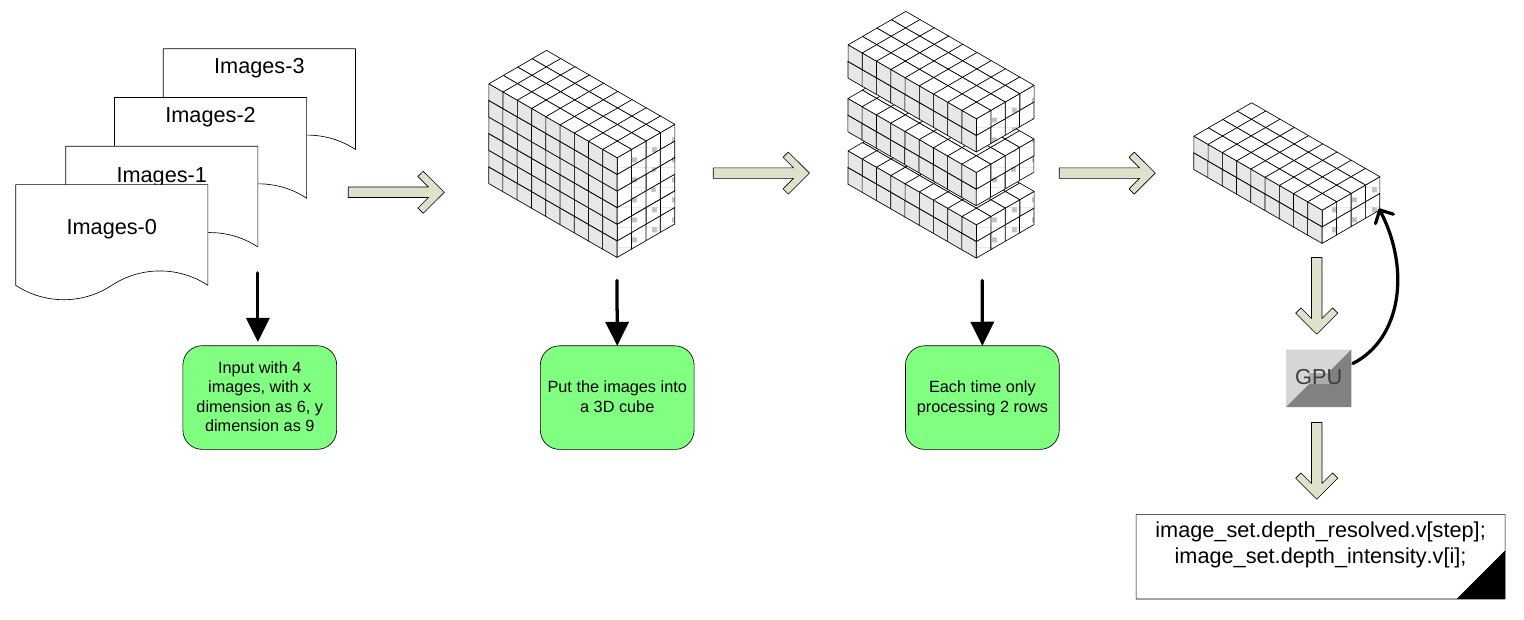}
    \caption{Cuda Program design}
    %\vspace*{3in}
    \label{fig:float}
\end{figure*}

To speed up the total time on reconstructing the image, we proposed a cuda program design using GPU to handle this problem. GPU is known for its high performance capability on computation intensive applications. For designing an efficient cuda program, several challenges which could decide the final program performance needs to be considered in advance. The challenges are listed as the following:
\begin{itemize}
\item The data structure which map each data element to a kernel thread.
  \item The communication time used to transfer data from CPU to GPU and vice versa needs to be minimized.
  \item The computation time spent on GPU for the cuda program needs to be minimized.
  \item Video card usually has limited memory.
\end{itemize}

For handling the first challenge, the special programming characteristic of cuda program needs to be taken into consideration. The speciality of Cuda programming characteristic is that it will launch multiple kernel thread on GPU side at same time. Each kernel thread is usually doing computation on each data element in the dataset and how to map the index of the element in the dataset with the index of the kernel thread in cuda will be a challenge which needs to be considered for designing the data structures.

\begin{figure}
    \centering
    \includegraphics[width = 0.95\linewidth]{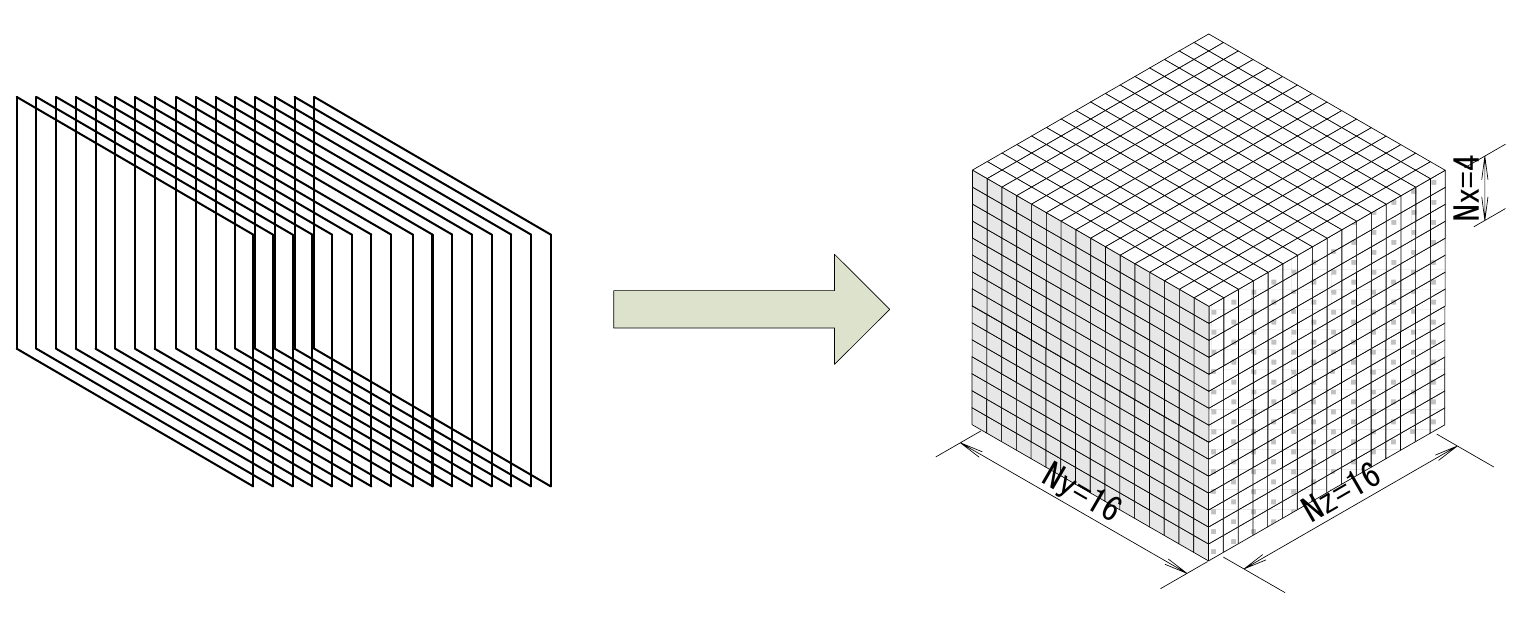}
    \caption{Data Input}
    \label{fig:3dcube}
\end{figure}

%the subscript is (x,y,z), where $0\leq x \leq m$, $0\leq y \leq n$, $0\leq z \leq p$,
As mentioned in the above section, the data input is $p$ 2D images and each image has $m$ rows and $n$ columns pixels as shown in Fig.~\ref{fig:3dcube}. Each pixel in the image must be mapped to a corresponding kernel thread. In order to map each kernel thread to a corresponding pixels in the image, we could either dynamically making a 3D array or a 1D array. Then map the pixel's subscription in the 3D or 1D array to kernel thread's id (x,y,z).

To choose the more efficient data structure from 3D array or 1D array, the challenge of minimizing communication time between CPU and GPU and the computation time spent on both side needs to be considered. In aspect of communication and computation time, these two data structures can incur big performance difference, under the assumption that all the arrays are created dynamically. From the the characteristic of cuda program design, the communication time is usually spent on transferring data between CPU and GPU. The computation time is spent on the depth reconstruction of each data element and index mapping which is used to map each data element index to each thread index.

The first method is to dynamically create a 3D array, with a 1D array of pointers pointing to a 2D arrays. The advantage of this methods is that the pixel can be accessed directly based on the array subscript (x,y,z). The disadvantage is that extra pointers need to be passed from CPU side to GPU side which incur extra transferring time.

The second method is to dynamically create a 1D array, with just one pointer pointing to the first pixels of the array. The advantage of this methods is that no extra pointers are created and passed to CPU, which saves communication time. However, the disadvantage part is that the array index needs to be changed back and forth from 3D index. Extra computation time is incurred on both the CPU side and GPU side.

\begin{figure}
    \centering
    \includegraphics[width = 0.95\linewidth]{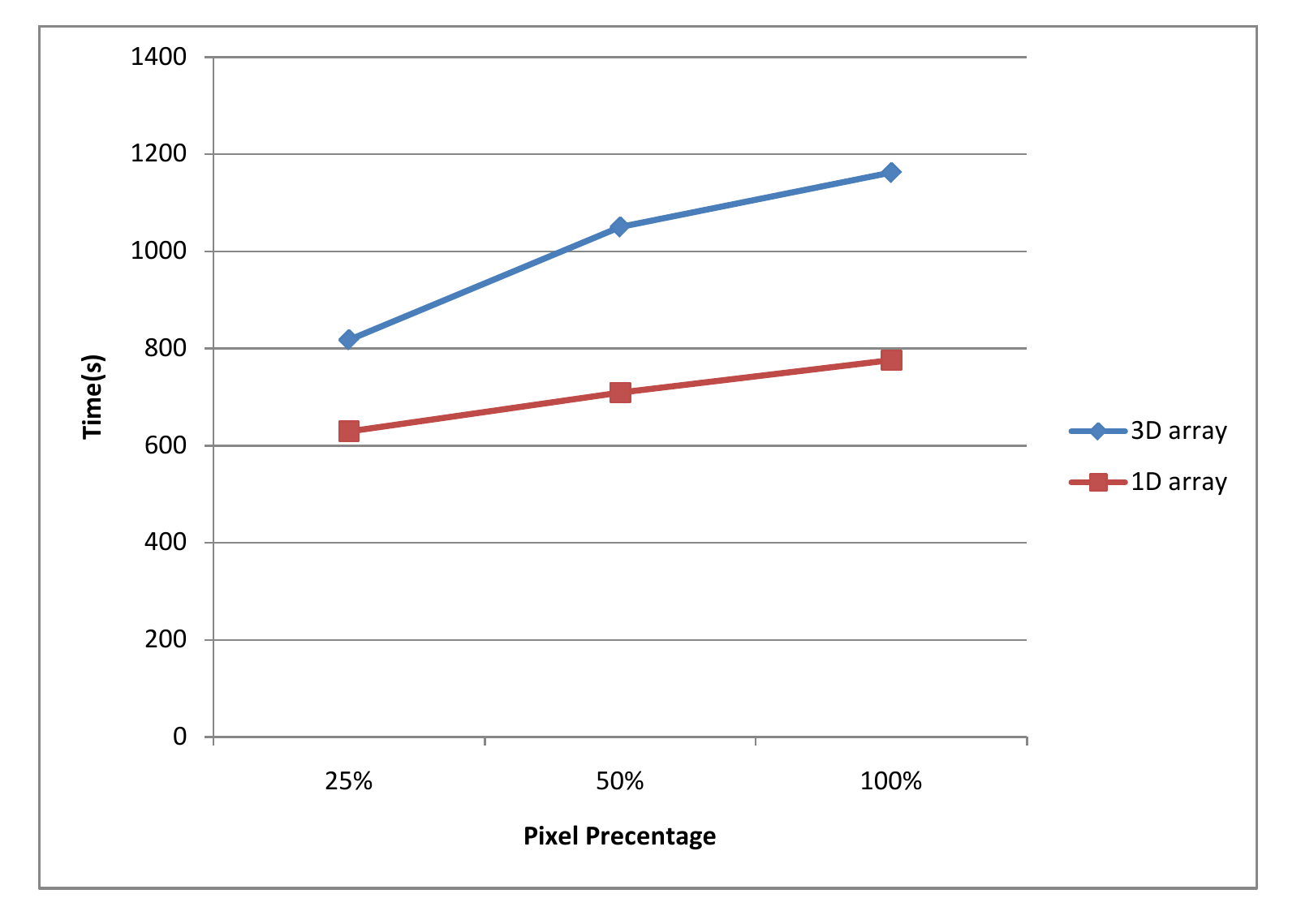}
    \caption{Performance Comparison Between 1D array and 3D array implementation}
    \label{fig:1d3dComp}
\end{figure}

The difference between these two methods is that the first one needs more communication time to copy extra array pointers from CPU to GPU, while the second method needs more computation time for changing the index back and forth between 3D index and 1D index. Thus there needs a consideration between computation and communication.  Communication time between CPU and GPU is usually a threshold for cuda program. Since we needs to pass extra pointers from CPU to GPU for 3D array design and GPU is known for its computation capability. It is easy to understand that 1D array would be better methods in terms of performance. Experiment is done on two different design for a 5G data set and the result is shown in Fig.~\ref{fig:1d3dComp}. It could seen from the result that 1D array design could save more time on the program computation. Thus, I implement the program using the first method of creating a 1D array as the data structure used in the CUDA program.

After determining the data structures, the limited memory on the video card needs to take into consideration. Since most GPU has limited memory and might not be able to handle all the data at one time. For example, the video card Tesla M2070 we used on our machine has maximum 6G memory. Other than the input data, the temporary data structures generated during the program has also to be count in. Thus, a best strategy is to divide the data input into several pieces and pass it to GPU. Take Fig.~\ref{fig:float} as example, there are 4 images with 6 rows and 9 columns as the data input, we put all the data into a 3D data cube. And each time we only pass 2 rows into GPU side, and return the result back to CPU side. This procedure will repeat for 3 times and the final result returned to CPU side will be put it back together. In Fig.~\ref{fig:dataflow}, detailed design is illustrated for each data set with 2 rows, 9 columns and 4 images.

\subsection{Detail Program Design}
\label{subsec:detail}

In order to make the GPU to process the data, cuda function cudaMemcpy is used to copy the data structures designed from CPU to GPU. And after the GPU processing, the result is copied back to CPU side via cudaMemcpy again. The parameter cudaMemcpyHostToDevice and cudaMemcpyDeviceToHost is used to specify which direction the data should be copied to.

\begin{figure}
    \includegraphics[width = 1.0\linewidth]{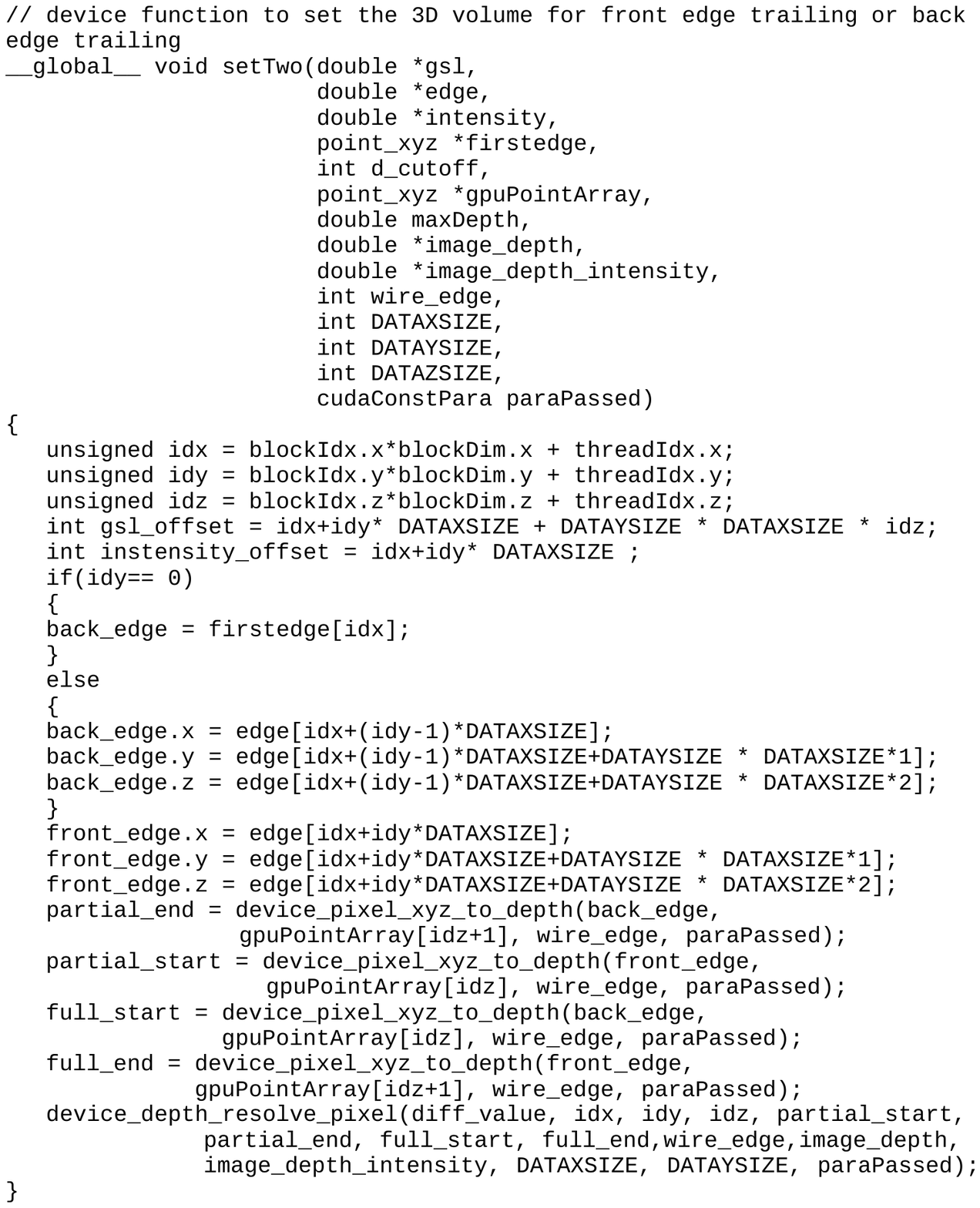}
    \caption{Cuda functions to start the GPU kernel}
    %\vspace*{3in}
    \label{fig:code1}
\end{figure}

Other than the computation part, the rest program, such as reading data from HDF5 files and writing result back to text files are still running on CPU. The start of the CUDA kernel function for computation is setTwo() function. The code slice for setTwo() function is shown in Fig.~\ref{fig:code1}. After passing the data from CPU to GPU, the kernel function started to do depth reconstruction on the data input.

Taking the example shown in Fig.~\ref{fig:float}, which has 4 images with 6 rows and 9 columns as the data input, we will illustrated the cuda kernel function in detail. Each time we only pass 2 rows out of 6 rows of data to GPU side. The columns is 9 and the number of images is 4, thus the total number of pixels would 72. Therefore we launch 72 kernel threads.

\begin{figure}
    \centering
    \includegraphics[width = 0.8\linewidth]{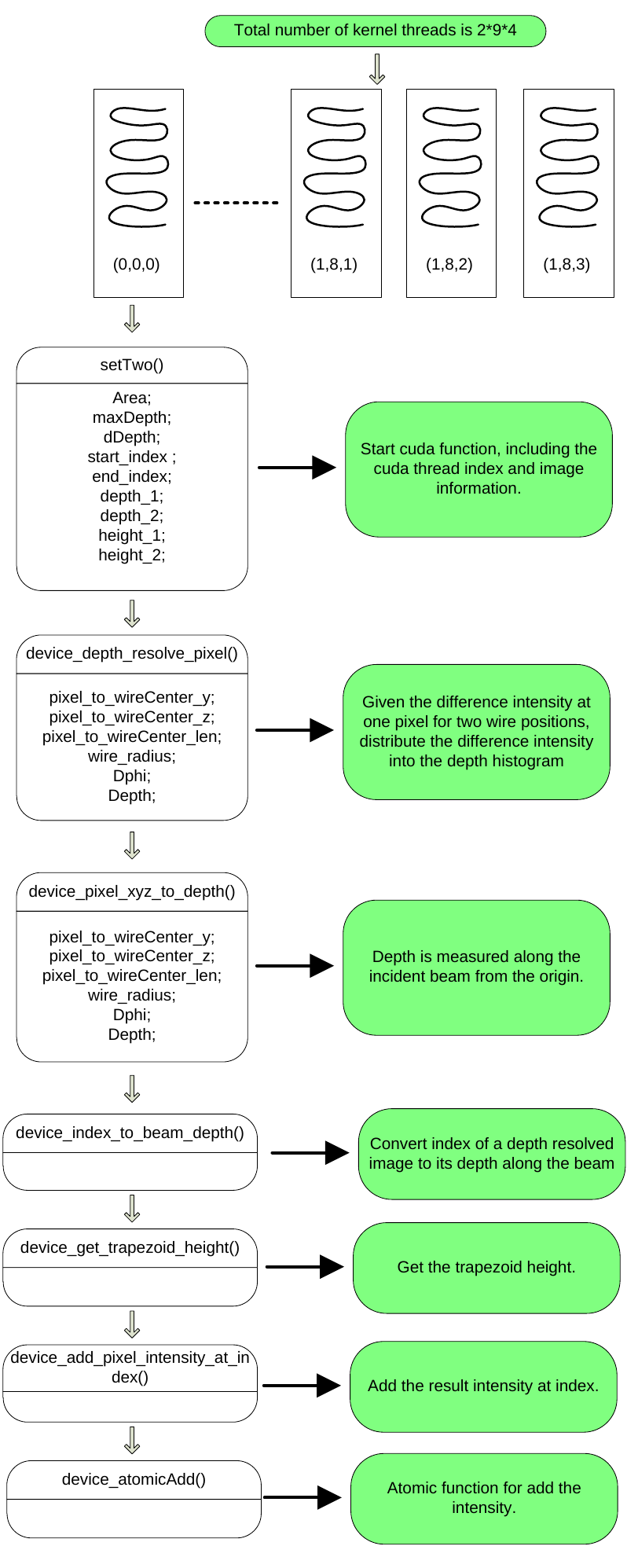}
    \caption{Program Flow}
    \label{fig:dataflow}
\end{figure}

As showed in Fig.~\ref{fig:dataflow}, total number of 72 kernel threads with the dimension (2,9,4) are launched for the data set. Each kernel thread will do computation on each pixel with corresponding subscript. For example, kernel thread (1,8,1) will be mapped to data (1,8,1) in the array. Each kernel thread start with function setTwo(), which handles the thread index mapping, and call the pixel reconstruction function. The pixel reconstruction functions process the difference intensity at each pixel and then add them together. And when adding all the intensity together, an atomic cuda function is implemented for multiple kernel threads writing to the same arrays. Since cuda atomic functions only support integer, a function for double-precision floating-point numbers is implemented using atomicAdd().

The detailed program flow for each kernel function is illustrate in Fig.~\ref{fig:dataflow}.

%\subsection{Cuda Program Design}
%\label{subsec:programdesign}

\section{Evaluation}
\label{sec:experiment}

\begin{figure}
    \centering
    \includegraphics[width = 0.95\linewidth]{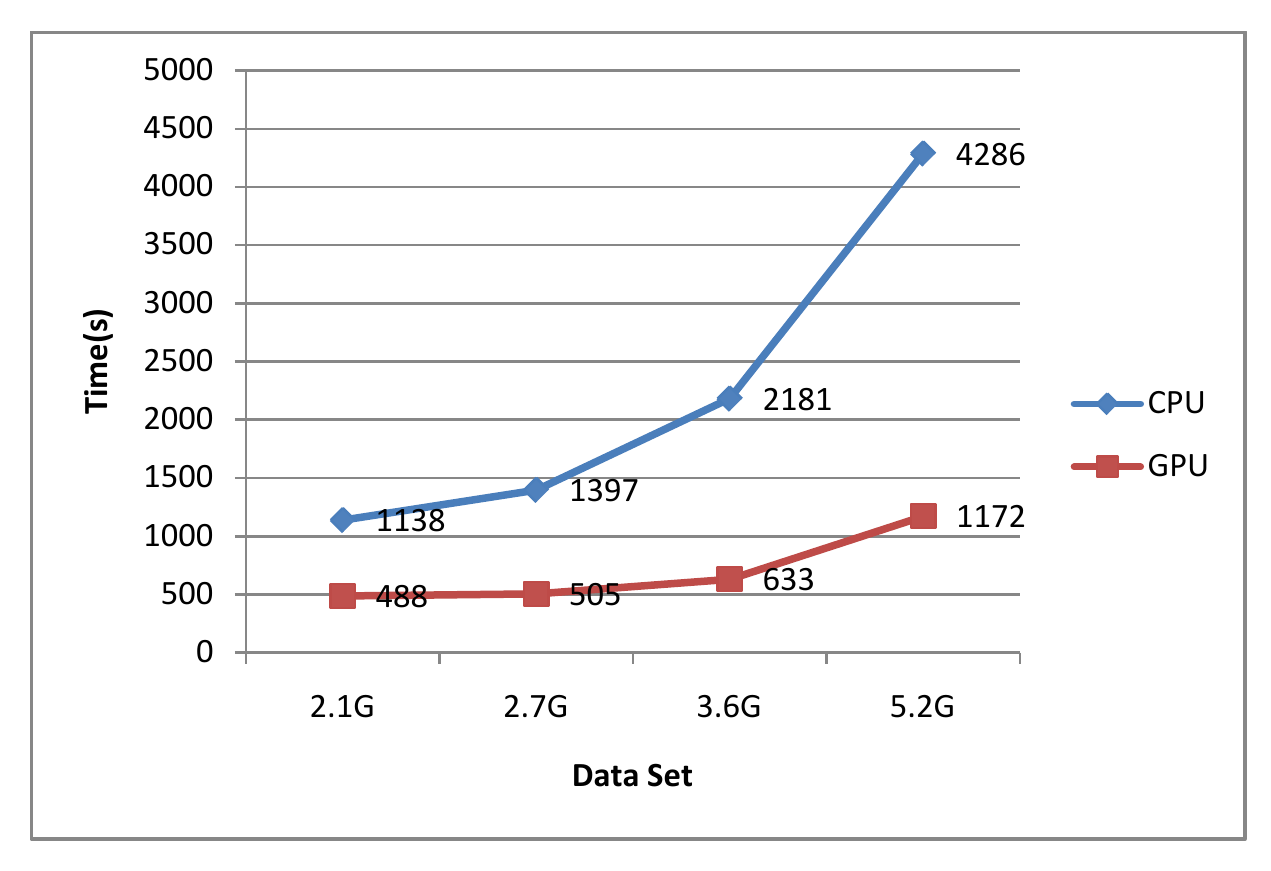}
    \caption{Performance Comparison between CPU and GPU for different data set}
    \label{fig:gpu01}
\end{figure}

In this section, experiments are performed on a GPU node running Linux operating system of the cluster to compare the performance for four different data sets. It has one Nvidia Tesla M2070 GPU installed on it, which has 6G memory. The number of maximum threads per block is 1024 and the maximum block dimensions and grid dimensions is $1024\times1024\times64$, $65535\times65535\times1$ separately. The node has a 4 core CPU (Xeon E5630) running at 2.53GHz and 32G of RAM. Makefiles for compiling the CUDA version has been added to the program and the GPU version of the program can be downloaded at https://subversion.xor.aps.anl.gov/DepthReconstruction. The experiment data input is hdf5 image acquired from the detector. The following figures are the comparison result of the final running time for different data sets and pixel percentage.

\subsection{Performance Test Result}
\label{subsec:result}

To compare the performance between CPU and GPU code, we run two experiments. The first is to change data sets from small size to big size. As showed in the Fig.~\ref{fig:gpu01}, we have four data sets with size 2.1G, 2.7G, 3.6G, 5.2G. The total memory used for CPU and GPU code are both 4G bytes. The final running time for CUDA design is just 25\% to 30\%  of the original CPU version. The GPU design of the image depth reconstruction outperforms CPU version in terms of performance.

\begin{figure}
    \centering
    \includegraphics[width = 0.95\linewidth]{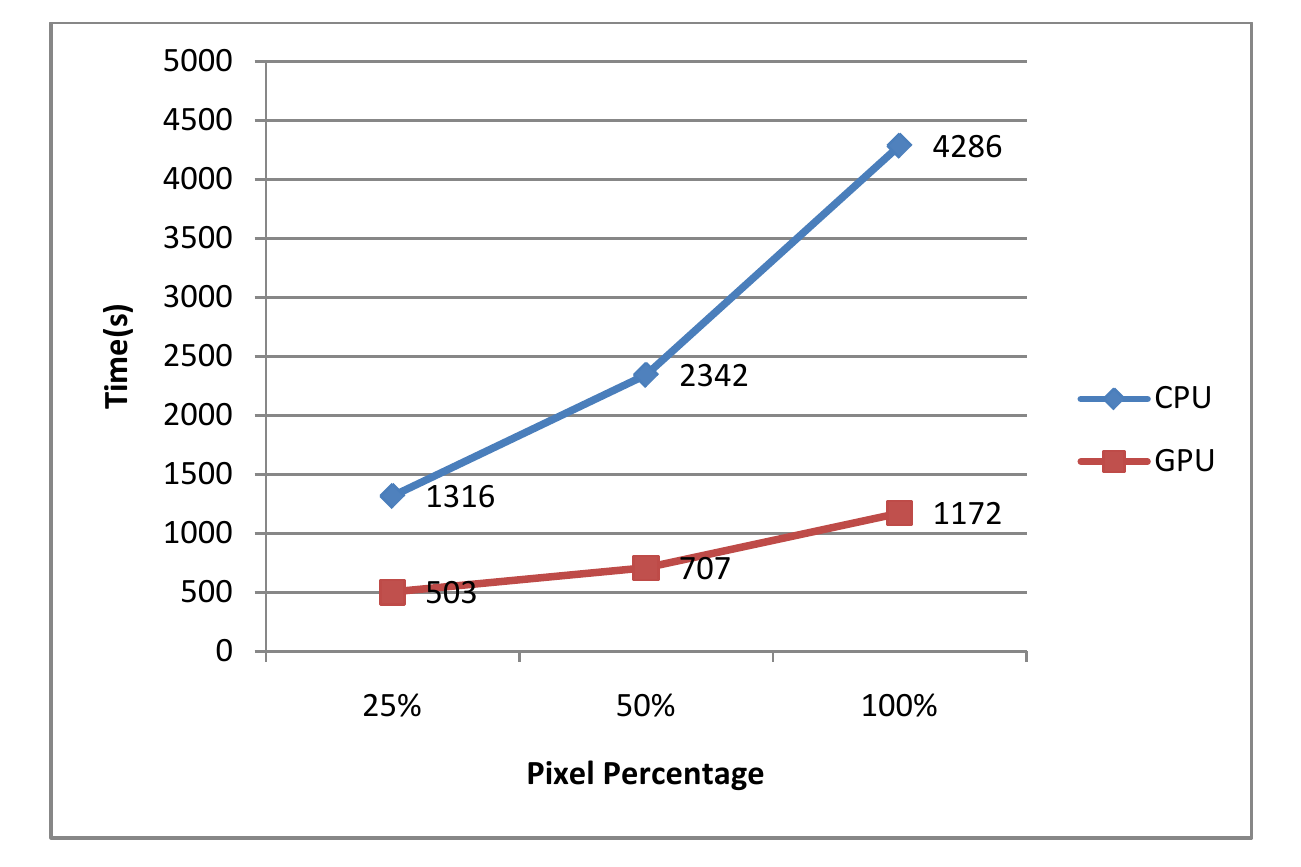}
    \caption{Performance Comparison between CPU and GPU for different pixel percentage}
    \label{fig:pixelpercentage}
\end{figure}

The second experiment is to change the percentage of the pixels in the data set. We change the pixel percentage to 25\%, 50\%, 100\%, and run two programs on the data set. The result is shown in Fig.~\ref{fig:pixelpercentage} and we can conclude that the more pixels we handle, the better performance we can get. When the pixel percentage is increasing, the more data is also transferred to GPU side which incurs more data communication time. While the time saved on computation still makes the total running time for GPU code less than CPU code.

Furthermore when the data sets become bigger from 2.1G to 5.2G as showed in Fig.~\ref{fig:gpu01}, we could see from the figure that the total running time for GPU did not change as much as CPU version. The conclusion can be got from the figure that our CUDA design did not just outperforms the CPU design in terms of performance, but also in scalability. Because GPU is usually doing very well in terms of computation intensive application. When we either changing the data size from 2.1G to 5.2G as showed in Fig.~\ref{fig:gpu01} or changing the pixel percentage from 25\% to 100\%, GPU needs to do more computation. This will enlarge the benefit that CPU can give it to the overall performance improvement, thus the explanation of the better scalability.

\section{Conclusion}
\label{sec:conclusion}
GPU with massively parallel structure has been proved more efficient than CPU~\cite{che2008performance} for processing large blocks of data in parallel. In this paper, we propose a GPU design methodology for the image depth reconstruction problem using CUDA. The test result shows that GPU design runs 3 or 4 times faster than the prior CPU design and thus gains a great performance improvement.  

\section*{Acknowledgment}
We thank Ruqing Xu from sector 34ID of Advanced Photon Source for providing HDF5 data and Sersted Roger for providing IT support.

\bibliographystyle{IEEEtran}
\bibliography{reference}

\end{document}